\documentclass
[prl,10pt,letterpaper,bibnotes,notitlepage,final,superscriptaddress,balancelastpage]{revtex4}

\usepackage{epsfig}
\usepackage{amsmath}
\usepackage{graphicx}
\usepackage{bm}
\usepackage{ulem}

\begin{document}

\title{How to create and detect N-dimensional entangled photons with an active phase hologram}

\author{Martin St\"utz}
\affiliation{Faculty of Physics, University of Vienna,
Boltzmanngasse 5, A--1090 Vienna, Austria}

\author{Simon Gr\"oblacher}
\email{simon.groeblacher@univie.ac.at} \affiliation{Faculty of
Physics, University of Vienna, Boltzmanngasse 5, A--1090 Vienna,
Austria} \affiliation{Institute for Quantum Optics and Quantum
Information (IQOQI), Austrian Academy of Sciences, Boltzmanngasse 3,
A--1090 Vienna, Austria}

\author{Thomas Jennewein}
\affiliation{Institute for Quantum Optics and Quantum Information
(IQOQI), Austrian Academy of Sciences, Boltzmanngasse 3, A--1090
Vienna, Austria}

\author{Anton Zeilinger}
\affiliation{Faculty of Physics, University of Vienna,
Boltzmanngasse 5, A--1090 Vienna, Austria} \affiliation{Institute
for Quantum Optics and Quantum Information (IQOQI), Austrian Academy
of Sciences, Boltzmanngasse 3, A--1090 Vienna, Austria}

\begin{abstract}
The experimental realization of multidimensional quantum states may
lead to unexplored and interesting physics, as well as advanced
quantum communication protocols. The orbital angular momentum of
photons is a well suitable discrete degree of freedom for
implementing high-dimensional quantum systems. The standard method
to generate and manipulate such photon modes is to use bulk and
fixed optics. Here the authors demonstrate the utilization of a spatial light
modulator to manipulate the orbital angular momentum of
entangled photons generated in spontaneous parametric
downconversion. They show that their setup allows them to realize photonic
entanglement of up to 21 dimensions, which in principle can be
extended to even larger dimensions~\cite{Stuetz2007}.
\end{abstract}

\maketitle

The ability to manipulate the light modes of single photons is a
crucial and necessary tool for any realization of high-dimensional
quantum optics experiments based on the orbital angular momentum of
photons~\cite{Mair2001,Vaziri2002a,Langford2004,Oemrawsingh2005,Marrucci2006}.
In addition, many other schemes of quantum optics experiments such
as quantum
teleportation~\cite{Bennett1993,Bouwmeester1997,Jennewein2001},
photonic quantum computing~\cite{Knill2001,Walther2005,Sanaka2004} and
higher-order photonic entanglement~\cite{Pan2000,Barreiro2005}
directly rely on photon interference, which requires to overlap
photon beams with high quality mode matching. The most common method
is to use bulk optical components such as lenses, mirrors and
microfabricated holograms in these setups. However, as such optical
components are static, each of them can only perform a predefined
manipulation of the photon beam. Programmable diffractive optics can
dramatically extend the capabilities of optical configurations by
allowing the generation of arbitrary and flexible light patterns in
real time~\cite{Jesacher2004,Gibson2004,Sinclair2004}.

Our work makes original use of a spatial light modulator (SLM) as an
active transformation in a quantum optics experiment for the manipulation of photons
coming from spontaneous parametric downconversion, which have
significantly less temporal and spatial coherence than photons from
a laser. The SLM manipulates one photon of a
photon pair entangled in its orbital angular momentum
(Laguerre-Gaussian functions) of the light mode and hence replaces
the fixed phase-singularity holograms used in preceding
experiments~\cite{Arlt1998,Mair2001,Vaziri2002a,Molina2004,Langford2004,Groeblacher2006,Inoue2006}.
We further demonstrate the possibility of generating 21-dimensional quantum states.

The SLM is an array of pixels (in our
case 1024 x 768 with a total size of 19.5 x 14.6~mm$^2$) acting as
individually tunable retardation wave plates which can imprint a
spatial phase modulation on a light beam. The SLM is made of nematic
liquid crystal pixels whose birefringence is controlled by an
external electric field. Our SLM is used in reflection mode, has a
refresh rate of 75~Hz and its diffraction efficiency into the first
order, which corresponds to the desired output, is approximately
60\%. For technical reasons we were limited by the nonideal
modulation depth of the SLM, which has a maximal phase shift of
$1.8\pi$ for our wavelength of $702$~nm~\cite{StuetzMaster}, a
reduced filling ratio of the pixels of 90\% and a reflectivity of
75\%.

 In order to generate the patterns to be applied onto
the SLM we implemented an algorithm on a computer in a \textsc{matlab}
environment. The function generating the output value for each pixel
on the SLM is
\begin{eqnarray}
\mbox{SLM\_Pixel}\left(\frac{x\cdot1024}{x_{\textrm{max}}-x_{\textrm{min}}},\frac{y\cdot768}{y_{\textrm{max}}-y_{\textrm{min}}}\right)=\nonumber\\
\;\;\;=\mbox{mod}\Bigl[(\mbox{angle}(\mbox{LG}(x-x_0,y-y_0,l,z,w_0,z_r))\label{LG_part}\\
\;\;\;+\frac{\pi\cdot10^3}{\lambda}\cdot f_{\textrm{SLM}}\cdot\left(\textrm{ast}\cdot(x-x_{l0})^2+(y-y_{l0})^2\right) \label{lense_part}\\
\;\;\;+x\cdot k_x+y\cdot k_y),2\pi\Bigr]\cdot\frac{256}{2\pi}
\label{mirror_part} \;\;\;,
\end{eqnarray}
where $\lambda$ is the wavelength of the photons impinging on the
SLM. This function produces a normalized 8 bits gray value ($0$ --
$255$) picture over a grid corresponding to the pixel positions of
the SLM. The standard range in meters for the values of $x$ and $y$
are \{0.0098,--0.0098\} and \{0.0073,--0.0073\}, respectively,
matching the size of the SLM.

 Expression~(\ref{LG_part}) of the function $\mbox{SLM\_Pixel}$ calculates the complex
amplitude $\mbox{LG}_{0,l}=\mbox{LG}(x-x_0,y-y_0,l,z,w_0,z_r)$ of
the desired Laguerre-Gaussian mode with the following parameters:
$z$ is the propagation direction of the light field, $w_0$ is the
beam waist, $z_r$ is the Rayleigh length and $x_0$ and $y_0$ are the
distance of the phase singularity from the origin. The function
"angle" is a \textsc{matlab} function which returns the complex phase angle
of the evaluated expression.

The index $l\in[\ldots,-1, 0, +1, +2,\ldots]$ is the azimuthal mode
index, with $2\pi l$ being the change in phase of a closed path
around the propagation axis. This phase change gives the Laguerre-Gaussian modes a helical wave front, where the angular momentum per
photon can only take integer multiples of $\hbar$~\cite{Allen1992}.
A phase hologram with $l\neq0$, imprints a phase $e^{-il\theta}$ on
the incoming beam, which results in a phase singularity. Such a
hologram with one singularity ($l=\pm1$) acts as a ladder operation
on the reflected light mode, effectively raising the $l$ index by
$1$ (hologram with $l=+1$) or lowering it by $1$ (hologram with
$l=-1$).

In addition, expression~(\ref{lense_part}) of the function
SLM\_Pixel allows the spatial light modulator to act as a Fresnel
lens. The parameter $f_{\textrm{SLM}}$ is the focal length of the lens term
in millimeters. Additionally, $x_{l0}$ and $y_{l0}$ are the distance of the
lens term from the origin in $x$ and $y$ direction, respectively.
The parameter ast is a weighting factor between the $x$ and the $y$ components of the lens term, which is used to compensate for any astigmatism. In the ideal case $\textrm{ast}=1$.

Furthermore, expression~(\ref{mirror_part}) of the function
SLM\_Pixel corresponds to a beam deflection, like a tilted mirror,
where the parameters $k_x$ and $k_y$ define an inclined plane and
are used to change the diffraction direction of the light beam, with
$k_x \approx 2\pi\theta_x/\lambda$ in the approximation of a small
diffraction angle $\theta_x$ (and analogous for $k_y$).

Effectively the SLM represents three adjustable diffractive optical
elements: a phase-singularity, a tunable lens and a tunable
mirror, which are readily usable in experiments for achieving mode
matching and beam pointing of a light beam.

\begin{figure}
\centerline{\includegraphics[width=0.6\textwidth]{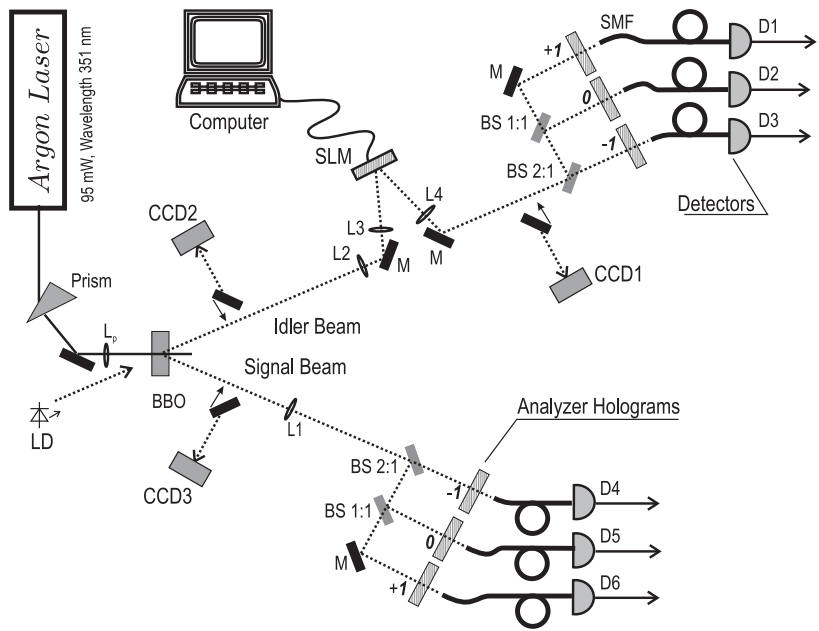}}\label{setup}
  \caption{Experimental setup to demonstrate the manipulation of entangled photons with a spatial light modulator. The optically nonlinear crystal (BBO) is pumped with an ultra-violet Ar$^{+}$ laser, which generates pairs of photons, entangled in their orbital angular momentum. In the idler beam the photons are controlled with a spatial light modulator and both signal and idler beam are analyzed with fixed phase holograms. The phase hologram of the spatial light modulator is actively controlled with a computer.}
\end{figure}

The experimental versatility of the SLM is demonstrated with the
manipulation of entangled photon pairs created by focusing an Ar$^+$
laser with a wavelength of 351~nm and a power of about $95$~mW into
a $\beta$-barium-borate crystal (BBO) with type-I mode matching (see Fig.~1 for the experimental setup). Due
to spontaneous parametric downconversion (SPDC), pairs of photons
entangled in their orbital angular momentum are
produced~\cite{Mair2001,Molina2005b,Huguenin2006}, called the signal
and the idler beam, both at a central wavelength of $702$~nm and a
bandwidth of 2~nm full width at half maximum. The photons in the signal beam pass a lens
$L_1$ ($f_1=250$~mm) to guarantee that they effectively couple into the
final fibers. In contrast, the idler beam is widened using lenses
$L_2$ ($f_2=-30$~mm) and $L_3$ ($f_3=100$~mm) to optimally
illuminate the SLM. With another lens $L_4$ ($f_3=750$~mm) the beam
is focused again after being reflected from the SLM. In order to
discriminate between the +1, -1 and 0 modes we use a probabilistic
mode analyzer (for a detailed description
see Ref.~\cite{Mair2001} and \cite{Vaziri2002a}).

In order to achieve a high pair coupling efficiency of both the
photons from the SPDC into the optical fibers, which is required for
maximal coincidence rates as well as perfect correlation in the
orbital angular momentum,  the signal and idler beams must have well
matched spatial mode functions. A striking advantage of the SLM over
fixed optics is the fact that now the idler beam can be easily matched to
the mode of the signal beam via tuning the lens term on the SLM. The
dependence of the spot size of the idler beam at the crystal on the
focal length $f_{\textrm{SLM}}$ of the SLM is measured by sending a laser
beam in reverse through the idler path, and measuring the mode
diameter with a charge coupled device (CCD) camera. The  focal length required to achieve
the same spot size for the signal and idler beams is found to be
$f_{SLM}=940$~mm.

Since the beam is reflected not perfectly orthogonal from the SLM,
the lens term applied to the SLM shows some astigmatism. Hence the
resulting mode of the reflected light beam will no longer be a pure
first order LG mode but a general superposition of Hermite-Gaussian
modes. This is compensated by tuning the parameter ast which puts an
astigmatism on the lens term, which also allows to perform
experiments of photons involving Hermite Gaussian
modes~\cite{MairPhD}. In our setup the ideal value to obtain clean
LG modes was $\textrm{ast}=1.029$ throughout the experiment [see Fig.~2(a)].

A further, important parameter is the position of the displacement
of the hologram in the $x$ direction $x_0$ and in the $y$ direction
$y_0$ with respect to the beam center. Tuning this parameter can be
used to obtain superpositions of different LG modes, as was done in
order to violate a Bell-type inequality~\cite{Vaziri2002a} and to
obtain a secret key in a quantum cryptography
scheme~\cite{Groeblacher2006}. Figure~2(b) shows the resulting mode
for different values of $x_0$ and $y_0$.

\begin{figure}
\centerline{
  \includegraphics[width=0.7\textwidth]{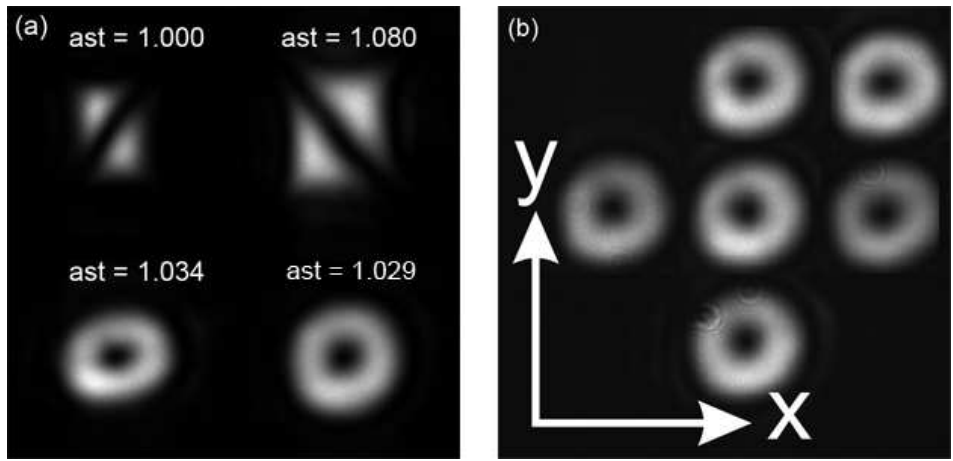}\label{lens}}
  \caption{Images (a) show how the astigmatism of the lens term is compensated
   with the parameter ast. For $\textrm{ast}=1.029$ an LG$_{0,1}$ is obtained (on the lower right), whereas for all other values general superpositions of Hermite-Gaussian modes are produced. (b) Pictures of the resulting mode for different values of $x_{l0}$ and $y_{l0}$, which determine the position of the lens term on the SLM. The camera is kept at the same position throughout the measurements.}
\end{figure}

The most important aspect of our experiment is to show that
entangled photons from downconversion can easily be manipulated
with the SLM device~\footnote{Our SLM-device is a LC--R 2500 from HoloEye
Photonics AG, Germany.}. We therefore place a CCD camera with a
long-pass filter (cutoff at $800$~nm) in the idler beam and apply
transformations between LG modes with different $l$ indices on the
SLM. As the majority of the downconversion photons are produced in
the LG$_{0,0}$ state, the intensity profiles on the CCD correspond
to the $l$ index of the transformation on the SLM. Figure~3(a) shows
images of the idler beam recorded for various values of $l$
($f_{\textrm{SLM}}=131$~mm and $k_y=2\times10^4$~rad/m). The mode with the highest
index shown is $l=10$, which can in principle be used for quantum
communication and information protocols with dimension of up to 21.
The transformations and therefore the accessible Hilbert spaces for
quantum optics experiments are in principle not limited.

\begin{figure}[t]
\centerline{
  \includegraphics[width=0.7\textwidth]{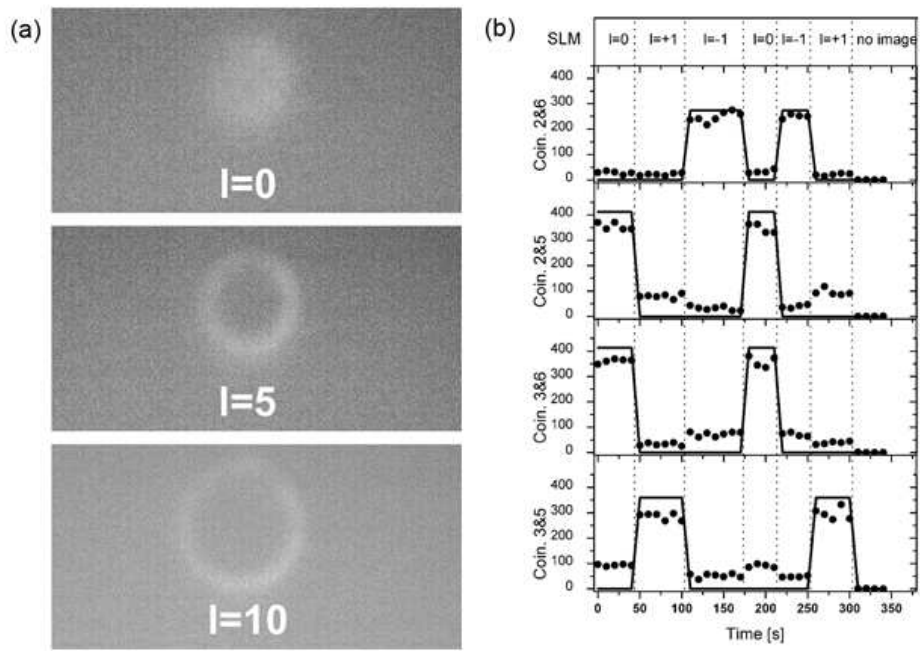}\label{coincpictures}}
  \caption{(a) Pictures of the downconverted light, transformed into higher-order modes by corresponding phase holograms on the SLM. (b) Coincidence counts of the downconverted beams per 10~s. With the SLM a transformation is performed and the correlations are observed via probabilistic mode analyzers. The individual settings are coupler 2, $l=0$; coupler 3, $l=-1$; coupler 5, $l=0$; and coupler 6, $l=+1$. The solid lines show the expected coincidences for different settings of the transformation hologram applied to the spatial light modulator (SLM). Only photon pairs with a total angular momentum $l_{\textrm{total}}=0$ are expected to show correlations due to the
  entangled nature of the generated two-photon state.}
\end{figure}

As a test of the SLM we analyze the perfect correlations of the
entangled photon pairs in their orbital angular momentum. The signal
beam is transformed with conventional quartz holograms and the idler
beam with the SLM scheme. These correlations are the crucial
ingredients for any quantum information processing scheme such as
quantum cryptography. The expected correlations for a maximally
entangled qutrit (a trinary quantum system; l=-1,0,+1) state for
four different couplers on the signal and idler side are shown in
Fig~3(b). Various phase holograms calculated for the desired
transformations were displayed on the SLM and the experimentally
obtained coincidences for the different couplers are also shown in
Fig.~3(b). The entangled character of the produced pairs can be
clearly seen and the ability of the spatial light modulator to
manipulate the pairs on the single photon level is proven. The
reason for the reduced visibility of the observed correlations in
the trinary quantum systems is mainly due to residual mismatching of
the two SPDC modes, as well as the nonperfect initial correlation
of the photon pairs produced in the downconversion process.

We have shown a scheme of manipulating entangled photons produced by
spontaneous parametric downconversion based on a spatial light
modulator. This scheme allows high flexibility for generating
arbitrary superpositions of orbital angular momentum states of
photons by tuning the position and the amplitude of
phase singularity. Transformations of the $l$ index of Laguerre-Gaussian modes up to 10 have been demonstrated, which paves the way
to experiments with 21-dimensional quantum systems. The SLM scheme
provides a fast and unique way of managing spatial modes of single
photons in quantum optics experiments.

\begin{acknowledgements}
The authors acknowledge financial support from the Austrian Science Fund
(FWF), the European Commission under the Integrated Project Qubit
Applications (QAP) funded by the IST directorate and the City of
Vienna.
\end{acknowledgements}

\end{document}